# PHOTONUCLEAR REACTIONS $^{nat}$Ni(γ,xn)$^{57}$Ni AND $^{nat}$Ni(γ,xn)$^{56}$Ni IN THE ENERGY RANGE $E_{γmax}$ = 35…94 MeV

*O. S. Deiev, I.S. Timchenko, S.M. Olejnik, S.M. Potin,*
*V.A. Kushnir, V.V. Mytrochenko, S.A. Perezhogin, V.A. Bocharov*
*National Science Center "Kharkov Institute of Physics and Technology", Kharkiv, Ukraine*
*E-mail: timchenko@kipt.kharkov.ua*

The total bremsstrahlung flux-averaged cross-sections ⟨σ($E_{γmax}$)⟩ for the photonuclear reactions $^{nat}$Ni(γ,xn)$^{57}$Ni and $^{nat}$Ni(γ,xn)$^{56}$Ni have been measured in the range of end-point energies $E_{γmax}$ = 35…94 MeV. The experiments were performed with the beam from the NSC KIPT electron linear accelerator LUE-40 with the use of the activation and off-line γ-ray spectrometric technique. The calculation of the flux-average cross-sections ⟨σ($E_{γmax}$)⟩$_{th}$ was carried out using the partial cross-section values σ(E) computed with the TALYS1.95 code and bremsstrahlung γ-flux calculated by GEANT4.9.2. The experimental results are compared with theoretical calculations and data from the literature. The obtained ⟨σ($E_{γmax}$)⟩ values expand the energy range of previously known experimental data.

PACS: 25.20.-x, 27.40.+z

## INTRODUCTION

Data on the cross-sections of photonuclear reactions with the yield of multiple nucleons are in demand in a variety of applications, for example, photoneutron generation, space nucleosynthesis, in the creation of nuclear installations based on subcritical systems (ADS) controlled by electron or proton accelerators, etc. At present, ADS systems are considered promising facilities for the disposal of radioactive waste from nuclear energy (transmutation of minor actinides [1] and burning of long-lived fission fragments), as well as potential sources for the production of electricity [2, 3]. Since such facilities operate in a subcritical mode, it is very important to know the flux of additional neutrons that are formed as a result of many-particle photoneutron reactions on the nuclei of structural materials (Al, Fe, Zr, Nb, etc.) inside of the ADS system. This will make it possible to more correctly estimate the total neutron flux in the core of a subcritical assembly, which, in turn, is related to the safety of operation of such nuclear installations.

Most of the data on the cross-sections of photonuclear reactions, important for many fields of science and technology, as well as for various data files (EXFOR, RIPL, ENDF, etc.), were obtained in experiments using quasi-monoenergetic annihilation photons in the energy range of the giant dipole resonance (GDR).

Photonuclear experiments with a large number of particles in the output channel require the use of high-intensity gamma-ray beams. Today, a good alternative is the use of high-energy electron bremsstrahlung passed through the converter. The use of bremsstrahlung significantly complicates the procedure for determining the characteristics of nuclear reactions. First of all, it is necessary to calculate the density of bremsstrahlung γ-quanta flux corresponding to the actual experimental conditions. In experiments, it is possible to measure the integral characteristics of the reactions. It required additional mathematical processing of the results. Nevertheless, bremsstrahlung of electrons is an important instrument of modern nuclear physics despite the difficulties in the determination of the reaction's cross-sections.

Interest in studies of photonuclear reactions with Ni isotopes is due to several reasons. Nickel is an important structural and surface coating material, used in accelerator nuclear technology and ADS [3]. Nickel is also used as a target material for the medical isotopes or radiopharmaceuticals, and radioactive sources.

In the GDR region, the study of photonuclear cross-sections on nickel isotopes was carried out in several works [4 - 7]. At higher energies, experimental studies of photonuclear reactions on stable nickel isotopes were carried out, for example, in a recent work [8]. In this work, using the γ-activation method, the flux-average photonuclear reaction cross-sections $^{nat}$Ni(γ,xn)$^{56,57}$Ni and $^{nat}$Ni(γ,pxn)$^{55,56,57,58}$Co were measured for end-point bremsstrahlung energies of 55, 59, 61, and 65 MeV.

This paper presents the total bremsstrahlung flux-averaged cross-sections ⟨σ($E_{γmax}$)⟩ for the $^{57}$Ni and $^{56}$Ni production in photonuclear reactions on natural nickel measured in the range of end-point bremsstrahlung energies $E_{γmax}$ = 35…94 MeV. The obtained data are compared both with the theoretical calculation performed using partial cross-sections σ(E) from the TALYS1.95 [9] code and with the data available in the literature [8].

## 1. EXPERIMENTAL PROCEDURE

### 1.1. EXPERIMENTAL PROCEDURE OF THE RESIDUAL γ-ACTIVITY OF THE IRRADIATED SAMPLE

Experimental studies of nickel photodisintegration cross-sections have been carried out through measurements of the residual γ-activity of the irradiated samples, which enabled one to obtain simultaneously the data from different channels of photonuclear reactions. This technique is well known and has been described in a variety of papers concerned with the investigation of multiparticle photonuclear reactions, e.g., on the nuclei $^{27}$Al [10], $^{93}$Nb [11 - 13], $^{181}$Ta [14, 15].

The schematic block diagram of the experiment is presented in Fig. 1. The γ-ray bremsstrahlung beam was generated by means of the NSC KIPT electron linac LUE-40 RDC "Accelerator" [16, 17]. Electrons of the initial energy $E_e$ were incident on the target-converter made from 1.05 mm thick natural tantalum plate, measuring 20 by 20 mm. To remove electrons from the bremsstrahlung γ-flux, a cylindrical aluminum absorber, 100 mm in diameter and 150 mm in length, was used. The targets of diameter 8 mm, placed in the aluminum capsule, were arranged behind the Al-absorber on the electron beam axis. For transporting the targets to the place of irradiation and back for induced activity registration, the pneumatic tube transfer system was used. After the irradiated targets are delivered to the measuring room, the samples are extracted from the aluminum capsule and are transferred one by one to the detector for the measurements. Taking into account the time of target delivery and extraction from the capsule, the cooling time for the sample under study took no more than 2 minutes.

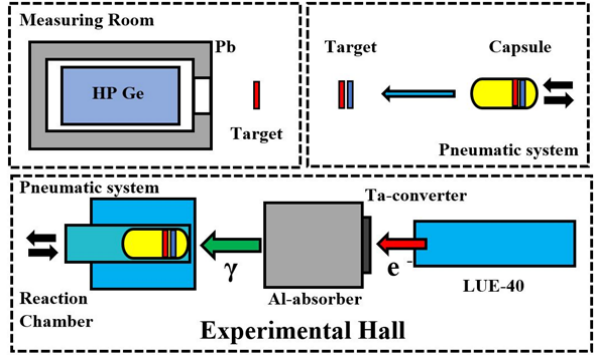

*Fig. 1. Schematic block diagram of the experiment. The upper part shows the measuring room, where the exposed target (red color) and the target-monitor (blue color) are extracted from the capsule and are arranged by turn before the HPGe detector for induced γ-activity measurements. The lower part shows the accelerator LUE-40, the Ta-converter, Al-absorber, exposure reaction chamber*

The induced gamma-activity of the irradiated targets was registered by the semiconductor HPGe detector Canberra GC-2018 with the resolutions of 0.8 and 1.8 keV (FWHM) for the energies $E_\gamma = 122$ and 1332 keV, respectively. Its efficiency was 20% relative to the NaI(Tl) detector, 3 inches in diameter and 3 inches in thickness. The absolute registration efficiency of the HPGe detector was calibrated with a standard set of gamma-ray radiation sources: $^{22}$Na, $^{60}$Co, $^{133}$Ba, $^{137}$Cs, $^{152}$Eu, $^{241}$Am.

The bremsstrahlung spectra of electrons were calculated by using the GEANT4.9.2 code [18] with due regard for the real geometry of the experiment, where consideration was given to spatial and energy distributions of the electron beam. The program code GEANT4.9.2 allows one to perform calculations taking properly into account all physical processes for the case of an amorphous target. In addition, the bremsstrahlung gamma fluxes were monitored by the yield of the $^{100}$Mo(γ,n)$^{99}$Mo reaction. For this purpose, the natural molybdenum target-witness, placed close by the target under study, was simultaneously exposed to radiation.

Natural Ni and Mo targets were used in the experiment. The isotopic composition of nickel is a mixture of five stable isotopes with an isotopic abundance: $^{58}$Ni – 0.68077, $^{60}$Ni – 0.26223, $^{61}$Ni – 0.0114, $^{62}$Ni – 0.03634, $^{64}$Ni – 0.00926. For $^{100}$Mo in our calculations, we have used the percentage value of isotope abundance equal to 9.63% (see ref. [18]). The admixture of other elements in the targets did not exceed 0.1% by weight.

In the experiment, the $^{nat}$Ni and $^{nat}$Mo samples were exposed to radiation at end-point bremsstrahlung energies $E_{\gamma max}$ ranging from 35 to 94 MeV with an energy step of ~ 5 MeV. The masses $^{nat}$Ni and $^{nat}$Mo targets were, respectively, ~ 80 and ~ 60 mg. The time of irradiation $t_{irr}$ and the time of residual γ-activity spectrum measurement $t_{meas}$ were 30 min for each energy $E_{\gamma max}$ value.

Table 1 lists the nuclear spectroscopic data of the radionuclide's reactions according to the data from [19]: $E_{th}$ denotes reaction thresholds; $T_{1/2}$ – the half-life period of the nuclei-products, $E_\gamma$ are the energies of the γ-lines under study and their intensities $I_\gamma$. To process the spectra and estimate the full absorption peak areas ΔA, the program InterSpec V.1.0.8 [20] was used.

There are two frames of the typical gamma-spectrum from reaction products of the nickel target in the $E_\gamma$ range from 100 to 600 keV and 1000 to 1500 keV shown in Fig. 2.

*Table 1*

*Nuclear spectroscopic data on the studied nuclei-products from the database [19]*

| Nuclide | $T_{1/2}$ | $E_\gamma$, keV | $I_\gamma$, % |
|---|---|---|---|
| $^{56}$Ni | 6.077±0.012 d | 158.38 | 98.8±0.1 |
| $^{57}$Ni | 35.60±0.06 h | 127.16 | 16.7±0.3 |
|  |  | 1377.63 | 81.7±1.6 |
| $^{100}$Mo | 65.94±0.01 h | 739.50 | 12.13±0.12 |

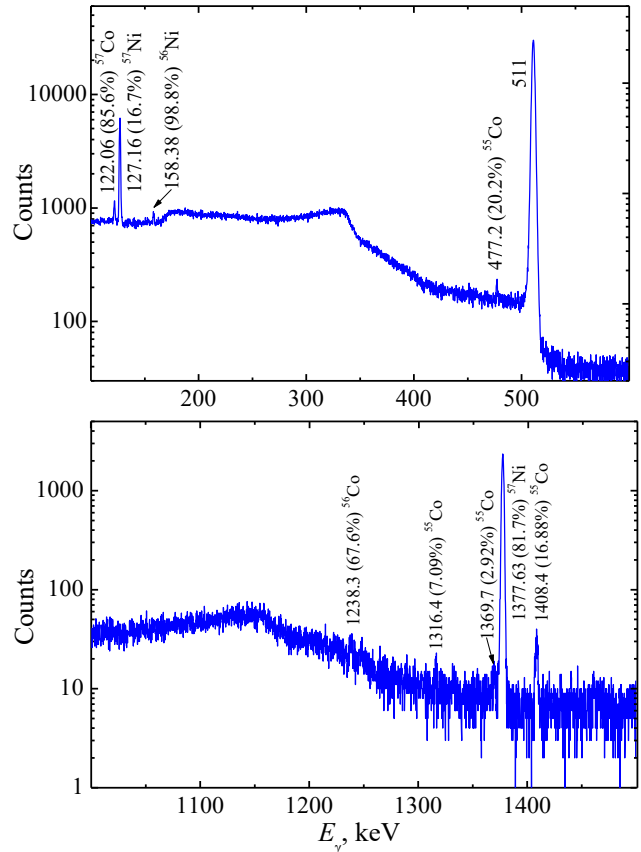

*Fig. 2. Fragments of the γ-radiation spectrum of $^{nat}$Ni, measured for the parameters: $t_{irr}$ and $t_{meas}$ = 1800 s, $t_{cool}$ = 2391 s, $E_{\gamma max}$ = 85.35 MeV, m = 82.105 mg. Spectrum fragments ranging from 100 to 600 keV and 1000 to 1500 keV are shown*

The bremsstrahlung gamma flux monitoring against the $^{100}$Mo($\gamma$,n)$^{99}$Mo reaction yield was performed by comparing the experimentally obtained flux-average cross-section $\langle\sigma(E_{\gamma max})\rangle$ values with the computation data. To determine the experimental $\langle\sigma(E_{\gamma max})\rangle$ values we have used the full absorption peak area $\Delta A$ value for the $E_\gamma = 739.50$ keV $\gamma$-line and the absolute intensity $I_\gamma = 12.13\%$ (see Table 1). The theoretical values of the average cross-section $\langle\sigma(E_{\gamma max})\rangle_{th}$ were calculated using the cross-sections $\sigma(E)$ in the TALYS1.95 code with the default options. The normalization (monitoring) factor $k_{monitor}$, derived from the ratios of $\langle\sigma(E_{\gamma max})\rangle_{th}$ to $\langle\sigma(E_{\gamma max})\rangle$, represents the deviation of the GEANT4.9.2-computed bremsstrahlung $\gamma$-flux from the real flux falling on the target. The determined $k_{monitor}$ values were used for normalizing the cross-sections for the studied photonuclear reactions. The monitoring procedure has been detailed in [11, 12].

### 1.2. EXPERIMENTAL ACCURACY OF AVERAGE CROSS-SECTIONS $\langle\sigma(E_{\gamma max})\rangle$

The uncertainty in measurements of experimental values of the flux-average cross-sections $\langle\sigma(E_{\gamma max})\rangle$ was determined as a quadratic sum of statistical and systematical errors. The statistical error in the observed $\gamma$-activity is mainly due to statistics calculation in the full absorption peak of the corresponding $\gamma$-line, which varies within 1 to 10%. This error varies depending on the $\gamma$-line intensity and the background conditions of spectrum measurements. The intensity of the $\gamma$-line depends on the detection efficiency $\varepsilon$, the half-life period $T_{1/2}$, and the absolute intensity $I_\gamma$. The background is generally governed by the contribution of the Compton scattering of quanta.

The systematical errors are due to the following uncertainties: 1) time of exposure and the electron current ~ 0.5%; 2) $\gamma$-ray registration efficiency of the detector ~ 2-3%, which is generally associated with the gamma radiation source error and the choice of the approximation curve; 3) the half-life period $T_{1/2}$ of the reaction products and the absolute intensity $I_\gamma$ of the analyzed $\gamma$-quanta were up to 4.7%, as is noted in Table 1 according to the data from [19]; 4) normalization of the experimental data to the yield of the monitoring reaction $^{100}$Mo($\gamma$,n)$^{99}$Mo made up to 5%. It should be noted that the systematic error in yield monitoring of the $^{100}$Mo($\gamma$,n)$^{99}$Mo reaction stems from three unavoidable errors, each running to ~ 1%. These are the unidentified isotopic composition of natural molybdenum ($^{100}$Mo isotope abundance equal to 9.63%), the uncertainty in the $\gamma$-line intensity used, $I_\gamma$, and the statistical error in the determination of the area under the normalizing $\gamma$-line peak.

Thus, the statistical and systematical errors are the variables, which differ for the investigated reactions. The total uncertainties of the experimental data are given in Figs. 4 and 5, and Tables 3 and 4 with experimental results.

### 1.3. CALCULATION FORMULAS FOR AVERAGE CROSS-SECTIONS

The cross-sections $\langle\sigma(E_{\gamma max})\rangle$ averaged over the bremsstrahlung $\gamma$-flux $W(E, E_{\gamma max})$ from the threshold $E_{th}$ of the reaction understudy to the end-point energy of the spectrum $E_{\gamma max}$ were calculated with the use of the theoretical cross-section $\sigma(E)$ values computed with the TALYS1.95 code [9], installed on Ubuntu20.04. The cross-sections $\sigma(E)$ for the $^{nat}$Ni($\gamma$,$xn$)$^{56}$Ni and $^{nat}$Ni($\gamma$,$xn$)$^{57}$Ni reactions, calculated by the TALYS1.95 code, are shown in Fig. 3.

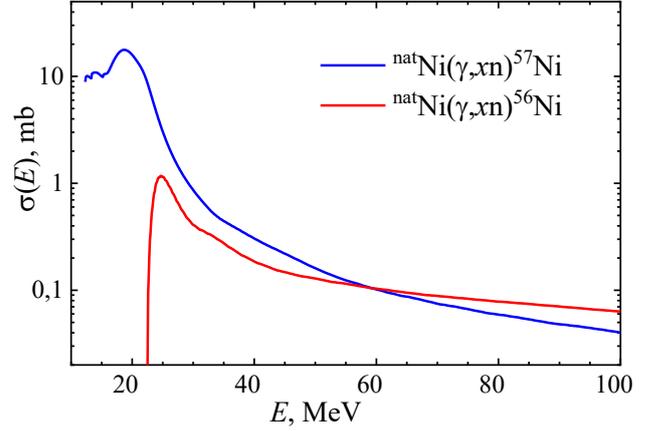

*Fig. 3. The cross-section $\sigma(E)$ for the reactions $^{nat}Ni(\gamma,xn)^{56}Ni$ and $^{nat}Ni(\gamma,xn)^{57}Ni$ (calculation by the TALYS1.95 code)*

The bremsstrahlung flux-averaged cross-section $\langle\sigma(E_{\gamma max})\rangle$ in a given energy interval was calculated by the formula:

$$\langle\sigma(E_{\gamma\max})\rangle = \frac{\int_{E_{th}}^{E_{\gamma\max}}\sigma(E)W(E,E_{\gamma\max})dE}{\int_{E_{th}}^{E_{\gamma\max}}W(E,E_{\gamma\max})dE}. \quad (1)$$

These evaluated average cross-sections were compared with the experimental values calculated by the formula:

$$\langle\sigma(E_{\gamma\max})\rangle = \frac{\lambda\Delta A}{\varepsilon N_x I_\gamma \Phi(E_{\gamma\max})(1-e^{-\lambda t_{irr}})e^{-\lambda t_{cool}}(1-e^{-\lambda t_{meas}})}, \quad (2)$$

where $\Delta A$ is the number of counts of $\gamma$-quanta in the full absorption peak, $\lambda$ is the decay constant ($\ln 2/T_{1/2}$), $N_x$ is the number of target atoms, $I_\gamma$ is the absolute intensity of the analyzed $\gamma$-quanta, $\varepsilon$ is the absolute detection efficiency for the analyzed $\gamma$-quanta energy, $\Phi = \int_{E_{th}}^{E_{\gamma\max}} W(E,E_{\gamma\max})dE$ is the integrated flux of bremsstrahlung quanta in the energy range from the reaction threshold $E_{th}$ up to $E_{\gamma max}$; $t_{irr}$, $t_{cool}$ and $t_{meas}$ are the irradiation time, cooling time and measurement time, respectively. A more detailed description of all the calculation procedures necessary for the determination of $\langle\sigma(E_{\gamma max})\rangle$ can be found in [11, 12].

The reaction yield is defined as:

$$Y(E_{\gamma\max}) = N_x \int_{E_{th}}^{E_{\gamma\max}}\sigma(E)W(E,E_{\gamma\max})dE. \quad (3)$$

This value is used in photonuclear experiments and is convenient for estimating the contributions of the reaction channels to the total reaction yield.

The normalized reaction yield contributions in the production of the $^{56,57}$Ni isotopes, calculated using the TALYS1.95 code with the default options, are presented in Table 2 for $E_{\gamma max}$ = 94 MeV. For lower energies $E_{\gamma max}$ contributions of the $^{58}$Ni($\gamma$,2n) and $^{58}$Ni($\gamma$,n) are close to 100%. The $^{58}$Ni($\gamma$,2n)$^{56}$Ni and $^{58}$Ni($\gamma$,n)$^{57}$Ni reactions are the dominant channels, respectively, in the $^{56}$Ni and $^{57}$Ni isotopes production on the natural nickel.

The normalized reaction yield contributions, given in Table 2, were used to estimate the experimental values of the flux-average cross-sections of the $^{58}$Ni($\gamma$,n)$^{57}$Ni and $^{58}$Ni($\gamma$,2n)$^{56}$Ni reactions, like in [21].

*Table 2*

*Normalized reaction yield contributions in the production of the $^{56,57}$Ni isotopes at $E_{\gamma max}$ = 94 MeV*

| Nuclide | Contributing reactions | $E_{th}$, MeV | Normalized yield, % |
|---|---|---|---|
| $^{56}$Ni | $^{58}$Ni($\gamma$,2n) | 22.47 | 99.75 |
| | $^{60}$Ni($\gamma$,4n) | 42.85 | 0.25 |
| $^{57}$Ni | $^{58}$Ni($\gamma$,n) | 12.22 | 99.79 |
| | $^{60}$Ni($\gamma$,3n) | 32.60 | 0.20 |

## 2. RESULTS AND DISCUSSIONS

The radionuclide $^{56}$Ni ($T_{1/2}$ = 6.077 d) is produced on all stable isotopes of natural nickel as a result of photonuclear reactions. To identify $^{56}$Ni it was used the $\gamma$-line with $E_\gamma$ = 158.38 keV, $I_\gamma$ = 98.8%. The obtained experimental data of the bremsstrahlung flux-averaged cross-sections $\langle\sigma(E_{\gamma max})\rangle$ for the $^{nat}$Ni($\gamma$,xn)$^{56}$Ni reaction are shown in Fig. 4. It can be seen that our experimental data were lower than the theoretical values $\langle\sigma(E_{\gamma max})\rangle_{th}$ calculated using the cross-sections $\sigma(E)$ from the TALYS1.95 code with the default options. At the same time, obtained $\langle\sigma(E_{\gamma max})\rangle$ are agree satisfactorily with data from [8].

The $^{56}$Ni nucleus in the investigated energy region is basically produced from the $^{58}$Ni isotopes by the $^{58}$Ni($\gamma$,2n) reaction, $E_{th}$ = 22.471 MeV. The $^{58}$Ni isotope contributed more than 99% to $^{56}$Ni production.

The radionuclide $^{57}$Ni ($T_{1/2}$ = 35.6 h) is produced on all stable isotopes of natural nickel as a result of photonuclear reactions. To identify $^{57}$Ni the $\gamma$-lines with $E_\gamma$ = 127.16 keV, $I_\gamma$ = 16.7% and $E_\gamma$ = 1377.63 keV, $I_\gamma$ = 81.7% were used. The experimental data of the bremsstrahlung flux-averaged cross-sections $\langle\sigma(E_{\gamma max})\rangle$ for the $^{nat}$Ni($\gamma$,xn)$^{57}$Ni are shown in Fig. 5. It can be seen that experimental data for both $\gamma$-lines are close to each other.

Such as in the case of the $^{nat}$Ni($\gamma$,xn)$^{56}$Ni reaction, the data for the $^{nat}$Ni($\gamma$,xn)$^{57}$Ni reaction are lower than the theoretical average cross-sections. The comparison of our $\langle\sigma(E_{\gamma max})\rangle$ with data [8] shows a significant discrepancy between the results.

The isotope $^{57}$Ni is basically produce from the $^{58}$Ni($\gamma$,n), $E_{th}$ = 12.218 MeV. The $^{58}$Ni isotope contributed more than 99% to $^{57}$Ni production (see Table 2).

The numerical values of the experimental total average cross-section data for the $^{nat}$Ni($\gamma$,xn)$^{56}$Ni and $^{nat}$Ni($\gamma$,xn)$^{57}$Ni reactions are presented in Tables 3 and 4, respectively. Also, the experimental $\langle\sigma(E_{\gamma max})\rangle$ values of the $^{58}$Ni($\gamma$,2n)$^{56}$Ni and $^{58}$Ni($\gamma$,n)$^{57}$Ni reactions with the taken into account the natural abundance of the $^{58}$Ni isotope and the normalized reaction yield contributions it can be found in Tables 3 and 4.

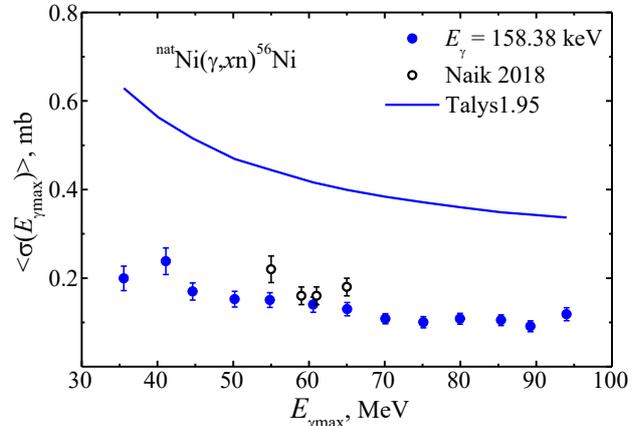

*Fig. 4. The bremsstrahlung flux-averaged cross-section $\langle\sigma(E_{\gamma max})\rangle$ for the reactions $^{nat}$Ni($\gamma$,xn)$^{56}$Ni: TALYS1.95 calculation – blue curve; experimental values: blue circles – our data $E_\gamma$ = 158.38 keV, black empty circles – Naik, 2018 [8]*

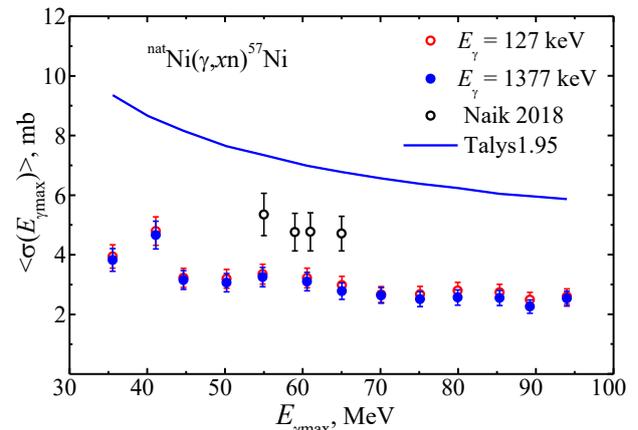

*Fig. 5. The bremsstrahlung flux-averaged cross-section $\langle\sigma(E_{\gamma max})\rangle$ for the reactions $^{nat}$Ni($\gamma$,xn)$^{57}$Ni: TALYS1.95 calculation – blue curve; experimental values: red empty circles – our data for $E_\gamma$ = 127.16 keV, blue circles – our data for $E_\gamma$ = 1377.63 keV, black empty circles – Naik, 2018 [8]*

## CONCLUSIONS

The total bremsstrahlung flux-averaged cross-sections $\langle\sigma(E_{\gamma max})\rangle$ for the photonuclear reactions $^{nat}$Ni($\gamma$,xn)$^{57}$Ni and $^{nat}$Ni($\gamma$,xn)$^{56}$Ni have been measured in the range of the end-point energies $E_{\gamma max}$ = 35…94 MeV. The experiments were performed with the beam from the electron linear accelerator LUE-40 NSC KIPT using of the activation and off-line $\gamma$-ray spectrometric technique. The calculations of flux-average cross-sections $\langle\sigma(E_{\gamma max})\rangle_{th}$ were carried out using the cross-section $\sigma(E)$ values computed with the TALYS1.95 code and bremsstrahlung flux calculated by GEANT4.9.2 code.

The experimental results are compared with theoretical calculations and literature data. Our data were lower than the theoretical $\langle\sigma(E_{\gamma max})\rangle_{th}$ values for both investigated reactions.

*Table 3*

*Experimental flux-average cross-section $\langle\sigma(E_{\gamma max})\rangle$ for the $^{nat}Ni(\gamma,xn)^{56}Ni$ and $^{58}Ni(\gamma,2n)^{56}Ni$ reactions*

| $E_{\gamma max}$, MeV | $^{nat}Ni(\gamma,xn)^{56}Ni$ | $^{58}Ni(\gamma,2n)^{56}Ni$ |
|---|---|---|
| 35.55 | 0.199 ± 0.028 | 0.293 ± 0.041 |
| 40.10 | 0.238 ± 0.030 | 0.350 ± 0.044 |
| 44.65 | 0.167 ± 0.019 | 0.250 ± 0.029 |
| 50.20 | 0.152 ± 0.018 | 0.224 ± 0.026 |
| 54.85 | 0.151 ± 0.017 | 0.221 ± 0.024 |
| 60.55 | 0.140 ± 0.017 | 0.205 ± 0.025 |
| 65.05 | 0.130 ± 0.015 | 0.191 ± 0.022 |
| 70.10 | 0.108 ± 0.011 | 0.159 ± 0.017 |
| 75.10 | 0.100 ± 0.013 | 0.147 ± 0.019 |
| 79.95 | 0.108 ± 0.012 | 0.159 ± 0.018 |
| 85.35 | 0.105 ± 0.012 | 0.154 ± 0.018 |
| 89.25 | 0.091 ± 0.012 | 0.134 ± 0.018 |
| 94.00 | 0.118 ± 0.015 | 0.174 ± 0.021 |

*Table 4*

*Experimental flux-average cross-section $\langle\sigma(E_{\gamma max})\rangle$ for the $^{nat}Ni(\gamma,xn)^{57}Ni$ and $^{58}Ni(\gamma,n)^{57}Ni$ reactions*

| $E_{\gamma max}$, MeV | $^{nat}Ni(\gamma,xn)^{57}Ni$ | $^{58}Ni(\gamma,n)^{57}Ni$ |
|---|---|---|
| 35.55 | 3.83 ± 0.38 | 5.62 ± 0.56 |
|  | *3.94 ± 0.39* | *5.79 ± 0.58* |
| 40.10 | 4.66 ± 0.47 | 6.85 ± 0.68 |
|  | *4.80 ± 0.48* | *7.04 ± 0.70* |
| 44.65 | 3.15 ± 0.31 | 4.63 ± 0.46 |
|  | *3.23 ± 0.32* | *4.74 ± 0.47* |
| 50.20 | 3.07 ± 0.31 | 4.50 ± 0.45 |
|  | *3.19 ± 0.32* | *4.69 ± 0.47* |
| 54.85 | 3.25 ± 0.32 | 4.77 ± 0.48 |
|  | *3.35 ± 0.33* | *4.92 ± 0.49* |
| 60.55 | 3.10 ± 0.31 | 4.55 ± 0.45 |
|  | *3.22 ± 0.33* | *4.72 ± 0.48* |
| 65.05 | 2.78 ± 0.28 | 4.08 ± 0.41 |
|  | *2.98 ± 0.30* | *4.37 ± 0.43* |
| 70.10 | 2.63 ± 0.26 | 3.86 ± 0.38 |
|  | *2.67 ± 0.27* | *3.92 ± 0.39* |
| 75.10 | 2.51 ± 0.25 | 3.68 ± 0.37 |
|  | *2.67 ± 0.27* | *3.92 ± 0.39* |
| 79.95 | 2.57 ± 0.26 | 3.77 ± 0.38 |
|  | *2.80 ± 0.28* | *4.10 ± 0.41* |
| 85.35 | 2.55 ± 0.25 | 3.74 ± 0.37 |
|  | *2.74 ± 0.27* | *4.02 ± 0.40* |
| 89.25 | 2.26 ± 0.23 | 3.31 ± 0.33 |
|  | *2.49 ± 0.25* | *3.65 ± 0.36* |
| 94.00 | 2.53 ± 0.25 | 3.71 ± 0.37 |
|  | *2.60 ± 0.26* | *3.81 ± 0.38* |

*The values $\langle\sigma(E_{\gamma max})\rangle$, obtained by the γ-lines with $E_\gamma$ = 1377.63 and 127.16 keV are marked in upright and in italics styles, respectively*

For the reaction $^{nat}Ni(\gamma,xn)^{56}Ni$, our experimental data agree satisfactorily with [8], while for the reaction $^{nat}Ni(\gamma,xn)^{57}Ni$ there is a significant discrepancy between the results.

It is shown that the $^{58}Ni(\gamma,n)^{57}Ni$ and $^{58}Ni(\gamma,2n)^{56}Ni$ reactions are the dominant channels, respectively, in the $^{56}Ni$ and $^{57}Ni$ isotopes production on the natural nickel.

The measured values $\langle\sigma(E_{\gamma max})\rangle$ expand the energy range of previously known experimental data and are useful in studying the mechanism of photonuclear reactions in the energy region above GDR.

## ACKNOWLEDGMENT

The authors would like to thank the staff of the linear electron accelerator LUE-40 NSC KIPT, Kharkiv, Ukraine, for their cooperation in the realization of the experiment.